\documentstyle[prl,aps,epsf,multicol]{revtex}
\epsfverbosetrue
\begin{document}
\draft


\title{Phase separation in the 2D Hubbard model: 
\\ a fixed-node quantum 
Monte Carlo study}

\author{A. C. Cosentini,$^{1,2}$
M. Capone,$^{1,2,3}$
L. Guidoni,$^{1,2,3}$ 
and G. B. Bachelet$^{1,2}$}
\address{$^{1}$Istituto Nazionale di Fisica della Materia (INFM), Italy}
\address{$^{2}$ Dipartimento di Fisica,
         Universit\`a di Roma La Sapienza, Piazzale Aldo Moro 2, I-00185
         Rome, Italy}
\address{$^{3}$International School for Advanced Studies 
(SISSA-ISAS), I-34014 Trieste, Italy}
\date{\today}

\maketitle

\begin{abstract}

Fixed-node Green's function Monte Carlo calculations have been 
performed for very large \hbox{$16\!\times\!16$} 2D Hubbard lattices, 
large interaction strengths $U\!=\!10, 20,$ and $40$, and many ($15 
\sim 20$) densities between empty and half filling.  The nodes were 
fixed by a simple Slater-Gutzwiller trial wavefunction.  For each 
value of $U$ we obtained a sequence of ground-state energies which is 
consistent with the possibility of a phase separation close to 
half-filling, with a hole density in the hole-rich phase which is a 
decreasing function of $U$.  The energies suffer, however, from a 
fixed-node bias: more accurate nodes are needed to confirm this 
picture.  Our extensive numerical results and their test against size, 
shell, shape and boundary condition effects also suggest that phase 
separation is quite a delicate issue, on which simulations based on 
smaller lattices than considered here are unlikely to give reliable 
predictions.

\end{abstract}


\begin{multicols}{2}


Strongly correlated electrons and holes are expected to play a key 
role in the high-$T_c$ superconductors.  Their possible instability 
towards phase separation (PS), initially believed to inhibit 
superconductivity, is attracting a lot of interest since a few 
different authors \cite{clc,emery,dagotto} have pointed out that such 
a tendency may in fact be intimately related to the high-$T_c$ 
superconductivity.  Long-range repulsive interactions may turn the PS 
instability into an incommensurate charge-density-wave (ICDW) 
instability, and the very existence of a quantum critical point 
associated to it may be a crucial ingredient of the superconducting 
transition \cite{clc2}.  PS and/or ICDW instabilities are related to a 
substantial reduction of the kinetic energy, which otherwise tends to 
stabilize uniformly distributed states; such a reduction is typical of 
strongly correlated electrons, both in real and model systems.

PS has been experimentally observed in La$_2$CuO$_{4+\delta} 
$\cite{jorgensen88,chou96}, where the oxygen ions can move: in the 
doping interval $0.01\leq \delta\leq0.06$ the compound separates into 
a nearly stoichiometric antiferromagnetic phase and a superconducting 
oxygen-rich phase.  In generic compounds, where charged ions cannot 
move, the possibility of a macroscopic PS is spoiled by the long-range 
Coulomb repulsion, and should lead to an incommensurate CDW 
instability \cite{tranquada}; here the identification of charge 
inhomogeneities with spoiled PS is less straightforward 
\cite{bianconi}.  On the theoretical side, evidence for PS has been 
suggested for various models of strongly correlated electrons, as the 
$t\!-\!J$ model \cite{emery2}, the three-band Hubbard model, the 
Hubbard-Holstein model and the Kondo model (see e.g.  Ref.  
\cite{clc2} and references therein).

Despite intensive studies, even for simple models there is no general 
agreement on the PS boundary: for the very popular $t\!-\!J$ model, PS 
is fully established only at large $J$, but at small $J$ (which 
unfortunately happens to be the physically relevant case) theoretical 
and numerical results are quite controversial.  Emery {\it et al.}'s 
\cite{emery2} theory that PS occurs at {\it any} value of $J$ in the 
$t\!-\!J$ model is confirmed by a recent numerical study by Hellberg 
and Manousakis \cite{hellberg}, but is in contrast with Dagotto {\it 
et al.}'s \cite{dagotto} exact numerical results on small clusters, 
suggesting no tendency toward PS for both the Hubbard model and the 
$t\!-\!J$ model below a critical value $J < J_c \sim t$, and with Shih 
{\it et al.}'s \cite{taipei} numerical results.  We also mention the 
recent suggestion by Gang Su \cite{gangsu}, according to which the 
Hubbard model does not show PS at {\it any} value of $U/t$ for any 
{\it finite} temperature, although it does not apply to ground-state 
properties.

PS is a thermodinamic instability associated to the
violation, in a given density range $n_1 < n <
n_2$, of the stability condition $\chi^{-1} = {\partial}^2 \!{\cal
E}/ \partial{n^2} > 0$, which requires the energy density ${\cal E}$
of an infinite electronic system to be a convex function of the
electron density $n$.
The system will therefore separate into two subsystems 
with electron densities $n_1$ and $n_2$. 
For the two-dimensional $t$-$J$ and Hubbard models, PS, 
if any, is expected to occur in a density range close to half filling 
($n \simeq 1$), and to yield a hole-rich phase with density $n_1<1$ 
and a hole-free phase with density $n_2=1$ \cite{emery2}.  In a truly 
infinite system such a PS would be associated with a vanishing inverse 
compressiblity $\chi^{-1}$ in the whole density range $n_1 < n < n_2$; 
in a finite system $\chi^{-1}$ may even become negative, because of 
surface effects.  So for finite systems it's preferable to pinpoint 
the PS using a Maxwell's construction (originally suggested by Emery, 
Kivelson and Lin \cite{emery2}, see also below).  But even such a 
procedure can give reliable results only for medium-large finite 
systems; really small systems (for which most numerical results have 
been up to now available) can attain so few and coarse densities, and 
suffer from so large finite-size errors, that their predictions on the 
relevant trends remains largely inconclusive.

Under these circumstances the availability of the fixed-node Green's 
function Monte Carlo (FNMC), a new and powerful numerical technique 
\cite{bemmel} which allows the study of (previously unfeasible) large 
lattice-fermion systems, provides us with a powerful tool to 
further investigate the Hubbard model.  Whether the $2D$ Hubbard 
hamiltonian, a prototype for interacting electrons with no long-range 
repulsion, shows any instability towards PS, is a very interesting 
open question.  A numerical study may also shed some indirect light on 
two related issues: the $t\!-\!J$ model in the physical region of 
small $J$, and the adequacy of the one-band Hubbard hamiltonian to 
catch an essential aspect of high-$T_c$ superconductors.


To evaluate the ground-state energy of the Hubbard hamiltonian
\begin{equation}
\label{hubbardham}
H = -t\sum_{\langle i,j \rangle\sigma}(c^{\dagger}_{i\sigma}c_{j\sigma} + h.c.)
+ U\sum_{i} n_{i\uparrow}n_{i\downarrow}
\end{equation}
we thus implemented the FNMC method recently proposed for lattice 
fermions by van Bemmel {\it et al.} \cite{bemmel,ceperley}, which has 
been used by Boninsegni for frustrated Heisenberg systems 
\cite{boninsegni} and by Gunnarsson {\it et al.} for 
orbitally-degenerate Hubbard models \cite{gunnarsson}.

The Green's function Monte Carlo, after a sufficiently long imaginary 
time, projects out the ground-state component of any initial 
wavefunction; apart from transient estimates, which for large systems 
appear to be hazardous unless the initial variational wavefunction is 
sufficiently close to the exact one, this method is therefore not 
directly usable for fermions in our Hubbard model (as well as any 
other model whose Green's function is not positive everywhere).  The 
FNMC \cite{bemmel,ceperley} replaces the true hamiltonian by an 
effective hamiltonian which confines the Monte Carlo random walk 
within a single nodal region (a region of the configuration space 
where the guiding wavefunction never changes sign), and, in analogy 
with the continuum case \cite{fixednode,upbound}, it provides an upper 
bound for the true ground-state energy \cite{ceperley}.

We also implemented the tecnique proposed in Ref.\cite{CASO98}, which 
allows us to reproduce with a relatively small fixed number ($100 \sim 
200$) of walkers equally accurate results as those obtained by means 
of standard MC runs of more than $2000$ walkers.

The variational wavefunctions we use to guide the random walks and to 
fix the nodes are the product of a Gutzwiller factor and two Slater 
determinants of single-particle, mean-field wavefunctions for up- and 
down-spin electrons.  The optimal Gutzwiller parameter and mean-field 
wavefunctions (whose only parameter is the staggered magnetization) 
were preliminarly obtained, for each $U$ and density, by variational 
Monte Carlo (VMC) runs.

A few representative variational and FNMC energies are shown in 
Table~\ref{table1} for the $4 \!\times\!  4$ Hubbbard lattice, for 
which exact results \cite{sorella} are available.  As expected, the 
VMC energy is always above the FNMC energy, which for these coupling 
strengths is slightly ($\sim 3\%$) above the exact energy.  For 
comparison we show the Constrained-Path Monte Carlo (CPMC) energies of 
Zhang {\it et al.} \cite{gubernatis}, which also include a larger $16 
\!  \times \!  16$ lattice (last row).  Especially at large $U$'s our 
results appear of comparable quality as theirs.  As far as the $4 
\!\times\!  4$ results are concerned, we notice that for $N_{e}=10$, 
which corresponds to a closed-shell configuration, both FNMC and CPMC 
are much closer to the exact energy than for $N_{e}=14$, which 
corresponds to an open-shell configuration.  This could be a serious 
problem when numerically studying the behavior of the energy as a 
function of the density; the results presented here fortunately show 
that, for lattices larger than $12\!\times\!12$, the shell effects 
become almost irrelevant.

\begin{figure}
\begin{center}
\epsfxsize=500 pt
\centerline{\epsfbox{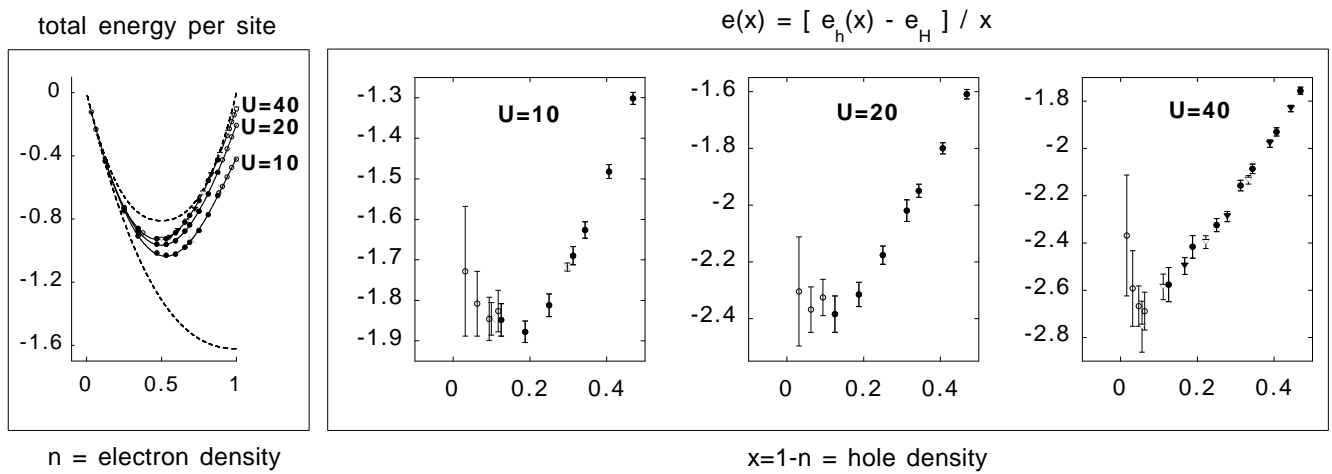}}
\end{center}

\caption{The first panel, to the very left, shows the ground-state 
energy per site (in units of the hopping parameter $t$) as a function 
of the electronic density, for a 2D Hubbard lattice of $N_{s} = 16 
\!\times\!  16 = 256$ sites with $U=10$ (lower), $20$ (middle), and 
$40$ (upper data).  Errors are smaller than the marker size.  Full 
markers correspond to closed shells and empty markers correspond to 
open shells.  The dashed curves correspond to two (analytically given) 
$U=0$ results: the fully spin-polarized case (upper curve), whose 
total energy per site is symmetric with respect to quarter filling, 
and the unpolarized case (lower curve), whose total energy per site is 
symmetric with respect to half filling.  Triangles (corresponding to a 
smaller $12 \!\times\!  12$ lattice and $U=40$) and crosses ($11\sqrt 
2\times 11\sqrt 2$ lattice and $U=10$) are shown for comparison (see 
text).  The second, third and fourth panel to the right contain plots 
of $e(x)$ vs.  $x$ (see text) for $U=10$, $20$, and $40$, 
respectively.  The data markers have the same meaning as in the first 
panel; obviously at small $x$ the error bar associated to $e(x)$, 
$\Delta{e(x)}=[\Delta{e}_{h}(x) +\Delta{e}_{h}(x\!=\!0)]/x$, becomes 
significant even if the statistical FNMC error $\Delta{e}_{h}(x)$ is 
tiny (see text).}

\label{fig1}
\end{figure}

To study the energy as a function of the electron density we have 
first tried out the less usual way of varying the density suggested by 
Ref.~\cite{hellberg} to avoid spurious Fermi-surface shape effects 
(keep the number of electrons $N_{e}$ fixed while the size of the 
underlying lattice is being varied), but discovered that either the 
number of electrons is really small (e.g.  $N_{e}\!=\!16$), and then 
artificial changes in the convexity of the curve may occur, or the 
system is large enough (e.g.  $12\!\times\!12$ lattices or larger), 
and then it doesn't matter how the density is being varied.  So for 
our systematic study (many densities and three $U$ values) we stick to 
a large $16 \!\times\!  16$ lattice ($N_{s}\!=\!256$ sites), and vary 
the number of electrons $N_{e}$ to yield electronic densities 
$n\!=\!N_{e}/N_{s}$ ranging from empty $n\!=\!0$ to half filling 
$n\!=\!1$.  In the first panel of Fig.~\ref{fig1} we show the 
electronic ground-state energy per site, obtained by FNMC runs as a 
function of the density\cite{apbc}.  Energies are in units of the 
hopping parameter $t$ throughout this paper; the statistical errors 
are smaller than the marker size, and thus are not visible here.  The 
calculated points are shown as full markers for closed shells, and as 
empty markers for open shells.  From Fig.~\ref{fig1} (see caption) it 
appears that the open-shell error, significant for a small 
$4\!\times\!  4$ lattice (see Table~\ref{table1}), becomes of the 
order of the statistical error (and thus negligible) for our large 
lattices \cite{shellstorte}.


At all densities our three sets of data for $U\!=\!10$ (lower), $20$ 
(middle), and $40$ (upper curve) are bracketed by the noninteracting 
unpolarized energy and the fully spin-polarized energy (both dashed in 
Fig.~\ref{fig1}), and display a smooth and reasonable behavior.  To 
evaluate the absolute accuracy of our results, we can rely on two 
exact limits: the low-density ($n\!\simeq\!0$) regime, where we expect 
${\cal E}=-4n$, and the half filled case ($n\!=\!1$),for which the 
strong-coupling expansion provides the correct large $U$ behavior: to 
leading order in $t/U$, the model maps onto an Heisenberg model, whose 
ground-state energy has been evaluated with great accuracy 
\cite{CASO98,SO93}.  We can also consider the next correction term 
$34.6 t^{4}/U^{3}$ \cite{Taka77}.  At low density our results are 
essentially exact; at half filling our error is small ($\sim\! 3\%$)
for $U=10$ but (as already seen in Table~\ref{table1})
it tends to grow with $U$: $\sim\! 9\%$ for $U=20$ and $\sim\! 11\%$
for $U\!=\!40$.  We have made sure (see markers other than dots in 
Fig.~\ref{fig1} and footnote \cite{apbc}) that such an energy 
discrepancy is not due to finite-size, shape, open-shell, and boundary 
condition effects; as far as systematic errors are concerned, we are 
thus left with the fixed-node approximation: as $U$ grows, more 
flexible trial wavefunctions than adopted here are required 
to obtain accurate nodes \cite{long}.

Keeping in mind the virtues and limitations of our numerical study, we 
can now turn to PS in the Hubbard model.  It has been shown in 
Ref.\cite{emery2} that the Maxwell construction is equivalent to 
study, as a function of the hole density $x \!=\!1\!-\!n$, the 
quantity $e(x) \!=\!  [e_{h}(x)\!-\!e_{H}]/x$, i.e.  the energy per 
hole $e_{h}(x)$ measured relative to its value at half filling 
$e_{H}=e_{h}(x \!=\!0)$.  For an infinite system, if the inverse 
compressibility $\chi^{-1}$ vanishes between some critical density 
$n_{c}<1$ and half filling $n=1$, then for $0 \leq x \leq x_{c}$ the 
function $e(x)$ is a constant, and the fingerprint of a PS is thus a 
horizontal plot of $e(x)$ below $x_{c}$.  For a finite system, 
instead, the PS fingerprint is a minimum of $e(x)$ at $x=x_{c}$ 
\cite{emery2}.  In some sense, $e(x)$ works like a magnifying lens of 
PS. It should be stressed that in a consistent definition of $e(x)$ 
the half-filling energy $e_{H}$ must be obtained as $e_{h}(x \!=\!0)$ 
from the same calculation as any other $e_{h}(x\neq 0)$ (in this work, 
from our FNMC).  If that's not the case, then $e(x)$ may tend to 
diverge near $x \!=\!0$, with the danger of artificially creating, 
rather than magnifying, the occurrence of PS. In the three right 
panels of Fig.~\ref{fig1} we find plots of $e(x)$ for $U=10$, $20$, 
and $40$; these values, as well as the associated error bars, are 
obtained from those of the first panel (original FNMC energies and 
tiny error bars).  Despite the error bars, a common trend is evident 
for all the calculated coupling strenghts: $e(x)$ has a positive slope 
for large hole densities, far from half-filling, but then it clearly 
changes slope below some small critical density $x_{c}$.  Such a 
minimum in $e(x)$ implies that, at least for the FNMC effective 
hamiltonian determined by our choice of wavefunction, PS occurs below 
$x=x_{c}$ \cite{size}.  Although a finer grid of hole densities would 
be required to locate with high precision the critical density $x_{c}$ 
as a function of $U$, we already see that $x_c$ decreases as $U$ is 
increased; this qualitatively agrees with the original predictions 
\cite{emery2} and with some previous calculations on the $t\!-\!J$ 
model at corresponding values of $J = 4t^2/U$ \cite{hellberg}.

In summary, our extensive FNMC numerical simulations of the Hubbard 
model for $16\!\times\!16$ two-dimensional lattices suggest PS for $U 
\gg t$.  If confirmed by further fixed-node simulations based on 
different nodes \cite{long} (and possibly even larger lattices 
\cite{size}), this result would imply that the $t\!-\!J$ model is also 
likely to show PS in the physically relevant regime $\!J < \!0.4$, and 
that even a single-band Hubbard model is sufficient to reproduce this 
physical tendency of high-$T_c$ superconductors.


We thank M. Boninsegni, S. Sorella, O. Gunnarsson, F. Becca, C. 
Lavalle, M. Grilli and C. Castellani for useful suggestions and 
discussions, and A. Filippetti for his precious help.  Special thanks 
are due to V.J. Emery and S.A. Kivelson, whose remarks were of key 
importance for the final discussion of our results, and to M. Calandra 
Bonaura and S. Sorella for making available to us the method of 
Ref.~\cite{CASO98} prior to publication and for many kind 
explanations.  GBB gratefully acknowledges partial support from the 
Italian National Research Council (CNR, Comitato Scienza e Tecnologia 
dell'Informazione, grants no.  96.02045.CT12 and 97.05081.CT12), the 
Italian Ministry for University, Research and Technology (MURST grant 
no.  9702265437) and INFM Commissione Calcolo.

\end{multicols}

\vspace{-0.5 truecm}

\begin{table}

\begin{center}
\begin{tabular}{cccccccccc}
& ${\rm size}$ & $N_{e}$ & 
 $n$ &  
 $U$ &
 ${\rm VMC}$  &
 ${\rm FNMC}$ &  
 ${\rm CPMC}$ &  
 ${\rm EXACT}$ & 
\\
\hline
& $ 4 \!\times\! 4 $ & $ 10 $ &
$ 0.625 $ & $ 4 $ & $ -1.211(2) $ & $ -1.220(2) $ & $ -1.2238(6) $ & $ 
-1.2238 $ &\\
& $ 4 \!\times\! 4 $ & $ 10 $ &
$ 0.625 $ & $ 8 $ & $ -1.066(2) $ & $ -1.086(2) $ & $ -1.0925(7) $ & $ 
-1.0944 $ &\\
& $ 4 \!\times\! 4 $ & $ 14 $ &
$ 0.875 $ & $ 8 $ & $ -0.681(2) $ & $ -0.720(2) $ & $ -0.728(3) $ & $ 
-0.742$ &\\
& $ 4 \!\times\! 4 $ & $ 14 $ &
$ 0.875 $ & $12 $ & $ -0.546(2) $ & $ -0.603(2) $ & $ -0.606(5) $ & $ 
-0.628$ &\\
&
$ 16 \!\times\! 16 $ & $ 202 $ &
$ 0.789 $ & $4 $ & $ -1.096(2) $ & $ -1.107(5) $ & $ -1.1193(3) $ & $ 
- $ &\\
\end{tabular}
\end{center}
\vspace{-0.2 truecm}
\caption{Ground-state energy per site (in units of the hopping 
parameter $t$) for a $4 \! \times \! 4$ Hubbard lattice and various
values of $U$. $N_{e}$ is the number of electrons and $n$ is the
corresponding average density. VMC: variational Monte Carlo, this work;
FNMC: Fixed-Node Green's function Monte Carlo, this work; CPMC: 
Constrained-Path Monte Carlo, Ref.~[21]; EXACT:
exact diagonalization results, Ref.~[20].(see text)}
\label{table1}
\end{table}

\begin{references}
\bibitem{apbc} Different boundary conditions yield energies within an 
error bar for $16 \!\times\!  16$ lattices and within at most two 
error bars for our smallest lattices; all results presented here refer 
to antiperiodic boundary conditions.
\bibitem{shellstorte} We have checked a few densities which
correspond to open shells in our usual $16\times 16$ lattice (empty 
dots in the first panel of Fig.~\ref{fig1}), but to closed shells in a 
$45^{o}$-rotated $11\sqrt 2\times 11\sqrt 2$ lattice (crosses in 
the first panel of Fig.~\ref{fig1}); the energy difference turns out 
to be negligible.
\end{references}
\end{document}